\documentclass[pre,showpacs,showkeys]{revtex4}

\usepackage{graphicx}
\usepackage{amssymb}

\newcommand{\beq}[0]{\begin{equation}}
\newcommand{\eeq}[0]{\end{equation}}
\newcommand{\e}{\varepsilon}
\newcommand{\thet}{\vartheta}
\newcommand{\la}{\langle}
\newcommand{\ra}{\rangle}
\newcommand{\ds}{\displaystyle}
\newcommand{\avh}{\la H_1 \ra}
\newcommand{\pa}{\partial}
\newcommand{\w}{\omega}
\newcommand{\ud}{{\mathrm d}}

\begin{document}

\title{Discrete Breathers in Hexagonal Dusty Plasma Lattices}

\author{V. Koukouloyannis}
\affiliation{Department of Physics, Section of Astrophysics, Astronomy and Mechanics, Aristotle University of Thessaloniki, 54124 Thessaloniki, Greece}
\affiliation{Department of Civil Engineering, Technological Educational Institute of Serres, 62124 Serres, Greece}

\author{I. Kourakis}
\affiliation{Centre for Plasma Physics, Queen's University Belfast, BT7 1NN Northern Ireland, UK}

\date{\today}

\begin{abstract}
The occurrence of single- or multisite localized vibrational modes, also called Discrete Breathers, in 2D hexagonal Dusty Plasma lattices is investigated. The system is described by a Klein-Gordon hexagonal lattice characterized by a  negative coupling parameter $\e$ in account of its inverse dispersive behavior. A theoretical analysis is performed in order to establish the possibility of existence of single- as well as three-site discrete breathers in such systems. The study is complemented by a numerical investigation based on experimentally provided potential forms. This investigation shows that a dusty plasma lattice can support single site discrete breathers while three-site in phase breathers could exist if specific conditions, about the inter-grain interaction strength, would hold. On the other hand, out of phase and vortex three site breathers cannot be supported since they are highly unstable.
\end{abstract}

\pacs{52.27.Lw, 52.35.Fp, 52.25.Vy, 63.20.Pw}

\keywords{Discrete Breathers, Multibreathers, Intrinsic Localized Modes,
Debye Crystals, Dust crystals, Dusty Plasma}

\maketitle

\section{Introduction}

Significant attention has been paid to highly localized time-periodic vibrating
motion in periodic lattices since roughly two decades ago. These modes [referring to \emph{discrete breathers} (DBs) and  earlier to \textit{intrinsic
localized modes} (ILMs)], owe their existence to lattice discreteness, in relation to the intrinsic nonlinearity
of the medium involved.
Following some initial phenomenological works in the late 80's (see e.g., in Refs. \cite{Page, Takeno, Dauxois, Kivshar, Campbell1}), a rigorous proof for
discrete breather existence was furnished independently by MacKay and Aubry \cite{MacKay1}
(later extended in \cite{Sepulchre}), using the notion of the {\it anticontinuous limit}, and by Flach
\cite{Flach1}, who used a {\it homoclinic orbit} approach.
Although single-site breathers possess the lion's share in this study, multi-site excitations (referred to as {\it multibreathers}) also exist and they are mentioned even since \cite{MacKay1} but they were studied more rigorously in later works. Some of the results in existence and stability of multibreathers in one dimensional Klein-Gordon chains can be found in Refs.\cite{ams, koukicht1, koukichtstability, arc} (also via the homoclinic orbit approach \cite{bountis, bountis1}). For the two-dimensional hexagonal case that interests us one can refer to \cite{koukmac} or, for instance,  \cite{PKevr1,pelkevfra,lawetal} for the DNLS case (also, see the references therein). Various studies have therefore been dedicated to different aspects
involved in the spontaneous formation, mobility and interaction of
DBs, both theoretically and experimentally; see e.g., in Refs.
\cite{Aubry,Flach2, MacKay2, Chaos, Campbell2, FlachGorbach} for a review.

On a separate physical playground, large ensembles of charged particles (plasmas) contaminated by massive heavily charged dust defects (\emph{dusty plasmas}, DP, or \emph{complex plasmas}) \cite{psbook, SVreview, SVbook, Morfill} host a wealth of novel linear and nonlinear collective effects, which are readily observed in laboratory
and space observations. Of particular importance is the occurrence of strongly coupled DP crystalline configurations,  typically formed and sustained in the sheath region above a horizontal negatively biased electrode in gas discharge (plasma) experiments \cite{psbook, Morfill}. These DP lattices are encountered in various configurations, including most commonly a superposition of levitated hexagonal two-dimensional (2D) layers \cite{Morfill}. One-dimensional (1D) configurations have also been created in the laboratory by making use of appropriate confining potentials \cite{TDLexper}.
Ultra-low-frequency modes (eigenfrequencies of only a few dozens of Hz) have been predicted, and later experimentally established, in the longitudinal (in-plane, acoustic mode), horizontal transverse (in-plane) and
vertical transverse (off-plane) directions \cite{psbook, SVreview, SVbook, Morfill}. The linear analysis of the latter (transverse, off-plane) degree of freedom has revealed the existence of a backward-wave character, as was theoretically predicted \cite{VSC1, VSC2} and later experimentally confirmed \cite{TDLexper}. Beyond 1D ``toy-lattice" phenomenology, studies of the same off-plane ``bending mode" have been carried out recently 2D DP crystals, establishing the inverse dispersive character of transverse dust lattice waves \cite{Vlad2, BFIKinprep}; those findings have successfully interpreted earlier numerical (molecular dynamics) \cite{Qiao} and experimental \cite{Samsonov2} results. It may be pointed out, for rigor, that the 2D picture bears substantially modified dispersion and group velocity characteristics, compared to the 1D case, as discussed extensively in the latter references.
The nonlinear modulation of continuum (as opposed to discrete, considered here) dust-lattice excitations in hexagonal crystals has been considered for the transverse mode in Ref. \cite{BFIKinprep}, and for horizontal (in-plane) motion in Ref \cite{BFIKPS2006}.

Here, we aim at focusing on the nonlinear aspects of the transverse (vertical) dust-lattice (TDL) mode, assuming the other two degrees of freedom to remain ``frozen" throughout this work. The necessary ingredients for discrete excitations to occur, namely \emph{nonlinearity} and lattice \emph{discreteness}, are present in the transverse dust lattice  mode \emph{par excellence}.
The sheath environment provides an on-site (``substrate") effective
potential (viz. the force is $F = -\partial \Phi/\partial z$) which may (for low density/pressure) be strongly \emph{anharmonic} near equilibrium, of the generic form:
\begin{equation}
\Phi(z) \approx \Phi(z_0) + \frac{1}{2} M \omega_0^2
(\delta z_n)^2 + \frac{1}{3} { \alpha} \, (\delta z_n)^3 +
\frac{1}{4}
 \beta \, (\delta z_n)^4 \, + {\cal O}[(\delta z_n)^5] \, .
 \label{Vzpol}
\end{equation}
The linear TDL eigenfrequency $\omega_0$ [in (\ref{Vzpol})], is typically as low as $\approx 20$ Hz \cite{VSC1,VSC2,IKIJBC}. The anharmonicity cofficients $\alpha$ and $\beta$ may be determined experimentally \cite{Ivlev2000,Zafiu,melzer}.
Sources of nonlinearity may include the electrostatic coupling anharmonicity \cite{IKPKSEPJDsols, IKIJBC, PhysScripta, BFIKPOP} and the coupling among different modes (either due to the interaction law \cite{Kompaneets} or as a purely geometric effect \cite{PhysScripta}). The anharmonicity (non-parabolic form) and the asymmetry (as in fact manifested by a strong cubic term) of the sheath potential form in (1) above is, as a matter of fact, associated with dust-grain charging and with the sheath collisionality, as theoretically shown from first principles in Refs. \cite{SV1} (note Fig. 8 therein) and \cite{SV2} (note Fig. 6 therein). Lattice \emph{discreteness}, on the other hand, is manifested via the weak interaction potential energy among neighboring grains, as compared to the energy ``stored" in an isolated (single-site) vibration. The former is measured by the characteristic coupling frequency
\beq
\omega_{T,0} = \biggl(\frac{Q^2}{M \lambda_D^3}\biggr)^{1/2} \biggl(\frac{1+\kappa}{\kappa^3}\biggr)^{1/2} e^{-\kappa/2} \, ,
\label{omegaT0Debye}
 \eeq
 (for Debye interactions) where $\kappa = a/\lambda_D$ is the lattice parameter (roughly of the order of unity), $Q$ is the electric charge of the dust grain and $\lambda_D$ is the Debye
 length, measuring the strength of Debye charge screening
 \cite{VSC1,VSC2,IKIJBC}. The latter (single-grain vibration energy) is measured by the linear transverse dust-lattice vibration eigenfrequency $\omega_0$ in (\ref{Vzpol}). Lattice discreteness is measured via the ratio
\beq
\delta = \omega_{T,0}^2/\omega_0^2 \, ,
\label{epsilondef}
\eeq
which
 acquires very small values, as suggested by experiments with 2D \cite{melzer} and earlier 1D \cite{Liu, Misawa} crystals. Dust lattices are therefore highly discrete systems, whose dynamics is located near the anticontinuous (uncoupled sites) limit  (since $\delta \ll 1$). All of these aspects can be inferred from earlier experimental studies \cite{Liu, Misawa}, although attention has not focused upon these aspects therein.

We have earlier suggested \cite{IKPKSPOPDB} that the vertical on-site potential anharmonicity, in combination with the
high discreteness of dust crystals, may enhance energy localization via the formation of discrete breather excitations. The occurrence of DBs in 1D dust crystals was investigated from first principles in Ref. \cite{IKVKPRE1}.
Our aim here is to extend those earlier results by elucidating the discrete dynamics of the transverse dust-lattice mode in 2D hexagonal crystals. A negative-valued ``spring constant'' is assumed where appropriate, in account of the inverse dispersion
inherent in TDL vibrations. Our results are tested for values from real plasma discharge experiments. The
analytical and numerical toolbox we used is described in full detail in earlier Refs. \cite{koukicht1,koukichtstability, kouk} and is therefore only briefly summarized here.

A number of comments are in row, to clarify our motivation and methodology. The scope of this article is two-fold. On one hand, we aim at presenting a generic investigation of the occurrence and stability of discrete lattice modes in hexagonal crystals, from first principles. On the other hand, the occurrence of discrete breather excitations associated with transverse dust lattice vibrations will be established through our work and lies in its very motivation.
This double-sided scope is reflected in the structure of the text, which first adopts a generic hexagonal lattice formulation and then focuses closer on dust-lattices. For the sake of reference and rigor, regarding the former component (generic hexagonal lattice dynamics), we may cite a number of existing studies of hexagonal lattice configurations.
Those earlier works have nevertheless focused on coupling (interaction) nonlinearity (e.g. via the Fermi-Pasta-Ulam paradigm, in fact disregarding on-site potential nonlinearity, a basic element in our case) \cite{Butt, Marin} or have employed the discrete nonlinear Schrodinger (DNLS) generic approach \cite{PKevr1, pelkevfra, lawetal}. Of particular interest is the study of localized structures in hexagonal photonic lattices; e.g. \cite{KivsharEtal1,KivsharEtal2,KivsharEtal3}. 
The nonlinear Klein-Gordon approach that we adopt here was considered in Ref. \cite{koukmac}, upon which we have relied in our methodology. Nevertheless, we here generalize by considering the occurrence of breathers in systems wherein linear modes obey an inverse dispersion law; for this purpose, we have adopted a formulation which leaves the sign of the coupling coefficient arbitrary (i.e., either positive or negative; read below). As regards the latter component we focus on (dusty plasma crystal transverse vibrations), earlier works \cite{BFIKinprep, BFIKPS2006} have adopted a continuum approximation, whose validity may be questioned in certain real experimental situations involving highly discrete DP crystal configurations.

The layout of the article proceeds as follows. Section \ref{HDhex} is devoted to the analytical formulation of a general model for hexagonal lattice dynamics. The existence and stability of discrete breather type excitations in such a lattice is investigated in Section \ref{HDhex-results} from a general point of view. Section  \ref{modelDP} is devoted to the analytical modelling of hexagonal dusty plasma crystals, in particular, while the generic results of
Sec. \ref{HDhex-results} are then explicitly applied to dusty plasma crystals in Section \ref{resultsDP}. Our results are finally summarized in the concluding Section.

\section{Hamiltonian Description of a Hexagonal lattice \label{HDhex}}

\begin{figure}[!ht]
\begin{center}
		\includegraphics[width=7cm]{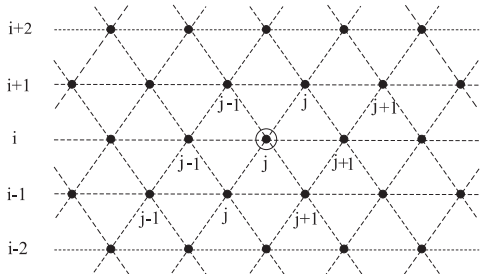}\vspace{-1cm}
	\caption{{\small The hexagonal lattice configuration is depicted, mounted with the site numbering employed in our model. The encircled oscillator is the ``central'' one.}}
	\label{lattice1}
	\end{center}
	\end{figure}

Consider a hexagonal lattice like the one shown in Fig.  \ref{lattice1}. Each site consists of a one-degree-of-freedom Hamiltonian oscillator with coupled with its six nearest neighbors
through a coupling constant $\e$.

The Hamiltonian of the full system is
\beq
\begin{array}{rl}
H=H_0+\e
H_1=&\ds\sum_{i,j=-\infty}^{\infty} \Big[ \frac{p_{ij}^2}{2}+V(x_{ij}) \Big] +\frac{1}{2}\,\frac{\e}{2}
\sum_{i,j=-\infty}^{\infty}\Big[(x_{ij}-x_{i-1,j})^2+(x_{ij}-x_{i-1,j+1})^2+\\+&\ds(x_{ij}-x_{i,j-1})^2
+(x_{ij}-x_{i,j+1})^2+(x_{ij}-x_{i+1,j-1})^2+(x_{ij}-x_{i+1,j})^2\Big] \, ,
\end{array}\label{ham}
\eeq
where the $i,j$ - indices denote the position of the oscillator in the lattice plane as it is shown in fig.\ref{lattice1}. The variable $x_{ij}$ denotes the displacement of the particle located at site ($i, j$) in the direction vertical to the lattice plane and $p_{ij}$ is the corresponding momentum. A factor $1/2$ is introduced in (\ref{ham}) in order for each site pair entering the summation to be considered only once.
Here it is assumed that the nonlinear on-site potential $V$ of the oscillator possesses a
stable
equilibrium at $(x,p)=(0,0)$, with $V''(0)=\w_{p}^2>0$.

\section{Existence and stability of discrete breathers in a hexagonal lattice \label{HDhex-results}}

\subsection{Existence and stability of single site breathers}

The existence of single site breathers can be investigated using the notion of the {\it anticontinuous limit} introduced in \cite{MacKay1}. This is constructed by taking $\e\rightarrow0$, i.e.\, by considering a chain of uncoupled oscillators. In this limit we consider one of the lattice oscillators, which we call ``central'', moving in a periodic orbit of period $T_b$ while the rest of them lying at rest at $(x,y)=(0,0)$. This state defines a trivially spatially localized and time periodic motion which is continued, for $|\e|$ small enough, to provide a single site breather of period $T_b$, if its period does not resonate with the phonon period, i.e.\, $T_b\neq mT_p=2\pi/\w_p, \forall m\in\mathbb{N}$ (e.g. \cite{MacKay1}). The profile of a sample single site DB can be seen in fig.\ref{single}.

The linear stability of a breather is determined by the
eigenvalues of the Floquet matrix $\lambda_i$ (called also Floquet multipliers) of the corresponding  periodic orbit, see e.g. \cite{Aubry,sepmacstab}. If all these multipliers lie on the unit circle of the complex plane then the periodic orbit is linearly stable, while if a multiplier lies outside the circle the orbit is unstable. Note that for every eigenvalue we also have its reciprocal and their complex conjugate(s), because of the Hamiltonian structure of the system (i.e.\ for every $\lambda_i$, there are also $\lambda_i^{-1}$, $\lambda_i^*$ and ${\lambda_i^*}^{-1}$). So, we cannot have just one eigenvalue outside the unit circle but only an even number of them. For
$\e=0$, the above mentioned multipliers lie in two complex conjugate bundles at
$e^{\pm i\w_pT_b}$, except of a pair of multipliers which lie at unity because of the phase degeneracy of the system. When the coupling switches on, for
$|\e|\neq0 \ll 1$, the breather is formed. Correspondingly, the multipliers of the non-central oscillators
move along the unit circle, without leaving it, because they are of the same Krein kind \cite{yac}, thus forming
the phonon band, while the pair which lies at unity remains intact. We keep increasing the value of $|\e|$ and the phonon band becomes wider. The DB, with the fixed period $T_b$, still exist and remains stable (changing thought its spatial profile) until the edge of the phonon band reaches $+1$.

Our purpose is to investigate if, for the (experimentally) given potentials, the DB can exist for the values of $\e$ also provided by experiments.

\subsection{Existence of three-site breathers}

\begin{figure}[!ht]
\begin{center}
		\includegraphics[width=7cm]{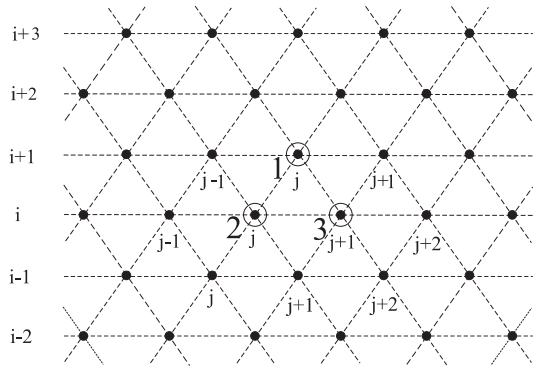}
	\caption{\small The hexagonal lattice, where the three ``central'' oscillators are shown.}
	\label{lattice2}
	\end{center}
	\end{figure}

We shall now consider the anticontinuous limit, where for $\e \rightarrow 0$ we consider three ``central''
oscillators (marked by circles in Fig.  \ref{lattice2}). These oscillators are moving in periodic orbits with
the same period $T_b=2\pi/\w$ but in arbitrary phases, while the rest lie at rest. As it is proven \cite{koukmac}, this state is continued for $\e\neq0$ and $|\e| \ll 1$, if specific conditions for the phase difference between the central oscillators hold in the anticontinuous limit.
The basic points of the procedure which is fully described in \cite{koukmac} are shown below.

Due to the phase degeneracy of the system, only the phase difference between the three central oscillators is significant in order to determine the specific periodic orbit for $\e=0$, which will be continued for $\e\neq0$ to provide the three-site breather. This way, a natural canonical transformation to the central oscillators is induced
\begin{equation}
\begin{array}{lr}
\begin{array}{rcl}
\theta&=&w_1\\
\phi_1&=&w_2-w_1\\
\phi_2&=&w_3-w_2
\end{array}$$&
\begin{array}{rcl}
A&=&J_1+J_2+J_3\\
I_1&=&J_2+J_3\\
I_2&=&J_3
\end{array}\end{array} \, ,
\end{equation}
where we have defined the action-angle variables $w, J$. Note, that $\phi_i$ denote the phase difference and since $\phi_3=w_1-w_3=-(\phi_1+\phi_2)$, there are only two independent $\phi_i$s, namely $\phi_1$ and $\phi_2$. Using these variables,  and adopting the notation of Ref. \cite{koukicht1}, the condition for existence of three-site breathers becomes
\beq\frac{\pa \avh}{\pa \phi_i}=0,\quad\left|\frac{\pa^2 \avh}{\pa \phi_i\phi_j}\right|\neq0\label{cond}\eeq
where
\begin{equation}
\avh=\frac{1}{T}\oint H_1 \ud t
\end{equation}
is the average value of $H_1$ along the unperturbed periodic orbit. It is assumed that two conditions hold, namely: a) non-resonance of the breather frequency with the phonon frequency $T_b\neq mT_p=m\frac{2\pi}{\w_p}$ and b) anharmonicity of single-site motion, viz. $\pa \w/\pa J\neq0$.

The motion of a single uncoupled oscillator of the lattice in the neighborhood of its stable equilibrium point can be described by a cosine Fourier series in the form
\beq x_i(t)=\sum_{n=0}^{\infty}A_n(J_i)\cos
nw_i=\sum_{n=0}^{\infty}A_n(J_i)\cos\left[ n(\w_i t+\thet_i)\right]\label{xt} \, . \eeq
Since we consider the central oscillators to move with the same frequency $\w_b$, the corresponding actions will have the same pertinent value $J_i=J$ and so will have the Fourrier coefficients $A_n(J_i)=A_n(J)=A_n$. Using (\ref{xt}), $\avh$ becomes
$$\la H_1\ra=-\frac{1}{2}\sum_{n=1}^{\infty}A_n^2\left\{\cos(n\phi_1)+\cos(n\phi_2)+\cos[n(\phi_1+\phi_2)] \right\},$$
and condition (\ref{cond}a) becomes
$$\frac{\partial\la
H_1\ra}{\partial\phi_i}=\frac{1}{2}\sum_{n=1}^{\infty}nA_n^2\left\{\sin
(n\phi_i)+\sin\left[ n(\phi_1+\phi_2)\right]\right\}=0.$$
This is satisfied for all choices of harmonic content if $\forall n \in
\mathbb{N}$,
$$\sin(n\phi_i)+\sin\left[ n(\phi_1+\phi_2)\right]=0, \qquad i=1,2.$$
At least three physically distinct solutions exist. The first one, with
\begin{equation}
\phi_{1,2}=0,
\end{equation}
corresponds to concerted motion of all three-sites oscillating in-phase, i.e. an {\it in-phase} three-site breather.
The second, with
\begin{equation}
\phi_{1,2}=\pi ,
\end{equation}
corresponds to an {\it out of phase} three-site breather. The final one, with
\begin{equation}
\phi_{1,2}=\frac{2\pi}{3},
\end{equation}
describes periodic motion of all three-sites in a vortex like ring, to be henceforth referred to as a \emph{vortex breather}.

The spatial profile of these three configurations can be found in figs. \ref{in},\ref{out},\ref{vortex}.

\subsection{Stability of three-site breathers}

As regards three-site breathers, for $\e=0$ there are the two bundles of Floquet multipliers at $e^{\pm i\w_pT_b}$, which correspond to the non central oscillators, and in addition there are six Floquet multipliers at unity, a pair for each of the central oscillators. As it is already mentioned in the case of single site breathers, for $|\e|\neq0$ the multipliers of the non-central oscillators spread along the unit circle, forming thus the phonon band. On the other hand, one pair of the multipliers that correspond to the central oscillators will always remain to unity, while the other two can move either along the unit circle or outside of it. If either one or both of these multiplier pairs leave the circle, the corresponding breather is unstable. On  the other hand, if they both leave unity along the unit circle, the corresponding breather is linearly stable for values of $|\e|$ small enough. For increasing values of $|\e|$ the breather remains linearly stable until the multipliers of the central oscillators reach the phonon band. Then, since these multipliers are of different Krein kind, the breather is destabilized through a Hamiltonian Hopf bifurcation and a complex quadruple of multipliers is generated outside the unit circle.

The Floquet multipliers of the central oscillators are given by $\lambda_i=e^{\sigma_iT}$ where $\sigma_i$ are the {\it characteristic exponents} of the periodic orbit which correspond to the DB, and correspond to eigenvalues of the stability matrix $E$ which in leading order of approximation is given by \cite{koukmac},
\beq E=\left(\begin{array}{cccc}
-\e\frac{\partial^2\la H_1\ra}{\partial\phi_1\partial
I_1}&-\e\frac{\partial^2\la H_1\ra}{\partial\phi_1\partial
I_2}&-\e\frac{\partial^2 \la
H_1\ra}{\partial\phi_1^2}&-\e\frac{\partial^2\la
H_1\ra}{\partial\phi_1\partial \phi_2}\\
-\e\frac{\partial^2\la H_1\ra}{\partial\phi_2\partial
I_1}&-\e\frac{\partial^2\la H_1\ra}{\partial\phi_2\partial
I_2}&-\e\frac{\partial^2 \la
H_1\ra}{\partial\phi_2\partial\phi_1}&-\e\frac{\partial^2\la
H_1\ra}{\partial\phi_2^2}\\
\frac{\partial^2H_0}{\partial I_1^2}+\e\frac{\partial^2\la
H_1\ra}{\partial I_1^2}&\frac{\partial^2H_0}{\partial I_1\pa
I_2}+\e\frac{\partial^2\la H_1\ra}{\partial I_1\pa
I_2}&\e\frac{\partial^2 \la H_1\ra}{\partial
I_1\partial\phi_1}&\e\frac{\partial^2 \la H_1\ra}{\partial
I_1\partial\phi_2}\\[8pt]
\frac{\partial^2H_0}{\partial I_2\pa I_1}+\e\frac{\partial^2\la
H_1\ra}{\partial I_2\pa I_1}&\frac{\partial^2H_0}{\pa
I_2^2}+\e\frac{\partial^2\la H_1\ra}{\pa I_2^2}&\e\frac{\partial^2 \la
H_1\ra}{\partial I_2\partial\phi_1}&\e\frac{\partial^2 \la
H_1\ra}{\partial I_2\partial\phi_2}
\end{array}\label{stab_mat}\right).\eeq
If all the eigenvalues of matrix (\ref{stab_mat}) lie on the imaginary axis, then the corresponding multipliers lie on the unit circle. If, in addition,
these eigenvalues are simple to first order in $\e$ (or have definite ``signature'', to
be explained later) then for small enough $\e$ the discrete breather is
linearly stable. On the other hand, if any of the eigenvalue pairs of $E$ is real then the corresponding DB is unstable.

The various components of the stability matrix $E$ are calculated as
\beq\begin{array}{cc}\ds\frac{\pa^2\avh}{\pa \phi_i\pa \phi_j}=&\left\{\begin{array}{ll}
\ds f(\phi)+f(2\phi)&j=i\\[8pt]
\ds  f(2\phi)&j\neq i
\end{array}\right.\end{array}\eeq
with
\beq
f(\phi)=\frac{1}{2}\sum_{n=1}^{\infty}n^2A_n^2\cos(n\phi),\eeq
and
\beq
\begin{array}{cc}\ds\frac{\pa^2H_0}{\pa I_i\pa I_{j}}=&\left\{\begin{array}{lcll}
\ds 2\frac{\pa^2H_0}{\pa J^2}&=&\ds2\frac{\ud \w}{\ud J}&, \quad j=i\\[8pt]
\ds-\frac{\pa^2H_0}{\pa J^2}&=&\ds-\frac{\ud \w}{\ud J}&, \quad j=i+1\\[8pt]
\end{array}\right.\end{array}
\eeq
The eigenvalues of $E$, i.e.\, the characteristic exponents of the DB, are to leading order of approximation,
\beq
\sigma_{1,2}=\pm\sqrt{-\e \frac{\ud\w}{\ud J} \, \bigl[2f(2\phi)+f(\phi)\bigr]}\, + {\cal O}(\e) , \qquad
\sigma_{3,4}=\pm\sqrt{-3\e\frac{\ud\w}{\ud J}\, f(\phi)}\, +{\cal O}(\e) \, . \eeq

We shall now investigate the behavior of these exponents for the 3 different modes introduced in the previous subsection. In doing so, we have to stress the significance of the product $P = \e \frac{\ud\w}{\ud J}$ (cf. the expressions above), which essentially determines the breather stability. We have chosen to keep the sign of the coupling strength $\e$ arbitrary, bearing in mind that inverse dispersive systems, such as transverse dust-lattice vibrations (to be discussed below), require for negative values of $\e$ to be considered in opposition to the usual Klein-Gordon configuration which is used for example for simulating the behavior of molecular crystals.

\paragraph{Stability of in-phase breathers.}

The characteristic exponents for the first class of time-reversible
solutions with $\phi_1=\phi_2=0$ are to leading order of approximation
\beq\sigma_{1,2,3,4}= \pm \sqrt{-3\e\frac{\ud \w}{\ud J}f(0)}\, + {\cal O}(\e) \, ,\label{s_in}\eeq
Since
$$f(0)=\frac{1}{2}\sum_{n=1}^{\infty}n^2A_n^2>0 \, , $$
the sign of the product $P = \e \frac{\ud\w}{\ud J}$ of the coupling strength with the anharmonicity, determines the linear stability of the
breather. If $P < 0$ the specific solution is unstable, since all eigenvalues $\sigma_{i}$ are real. On the other hand, for $P >0$, the leading order calculation suggests linear stability (imaginary $\sigma_{i}$). Because of the double multiplicity of $\sigma_i$ one could have expected that a splitting of the eigenvalues is possible, forming this way a complex quadruple which leads to instability. This, however, cannot happen due to symplectic signature reasons (the corresponding quadratic form is definite). This issue is thoroughly investigated in Refs. \cite{koukmac,macmei}.

Note that, the above stability results, as well as the ones that follow, apply both for $\e>0$ (which corresponds to a classical nonlinear Klein-Gordon lattice) and for $\e<0$ (which corresponds to Dusty Plasma Crystals).

\paragraph{Stability of out-of-phase breathers.}

For the second class of solutions, corresponding to $\phi_i=\pi$, the
eigenvalues of the stability matrix $E$ are
\beq\sigma_{1,2}=\pm\sqrt{-\e\frac{\ud\w}{\ud J} \, [2f(0)+f(\pi)]}\, +{\cal O}(\e) ,\qquad
\sigma_{3,4}=\pm\sqrt{-3\e\frac{\ud\w}{\ud J}\, f(\pi)}\, + {\cal O}(\e). \label{s_out}\eeq
Although the sign of the quantity
$$f(\pi)=\frac{1}{2}\sum_{n=1}^{\infty}(-1)^nn^2A_n^2 \, ,$$
is not prescribed in general, one concludes from the above that
\beq
2f(0)+f(\pi)>0 \, .
\eeq
Therefore, in order to have both pairs of eigenvalues lying on the imaginary axis, i.e.
for this class of solutions to be linearly stable, we shall require both conditions $f(\pi)>0$ and $P=\e
\frac{\ud\w}{\ud J}>0$ to be satisfied.

Note that, in most cases we have $f(\pi)<0$, since the first term -- which is usually dominant -- is a negative term. Therefore, considering the exponential decay of Fourier coefficients of smooth functions, the rest of the terms usually do not change the negative sign of the sum. If this fact holds it means that the quantities $f(\pi)$ and $2f(0)-f(\pi)$ will have opposite signs, so one of the $\sigma_i$ pairs will be real, which leads to instability.

\paragraph{Stability of vortex breathers.}

The eigenvalues of the matrix $E$ for the vortex configuration are
\beq
\sigma_{1,2,3,4}=\pm\sqrt{-3\e\frac{\ud\w}{\ud J}\, f_1} \, +\cal{O}(\e)\label{s_vortex} \, ,
\eeq
with
\beq f_1 = f\left(\frac{2\pi}{3}\right)=f\left(\frac{4\pi}{3}\right)=\sum_{n=1}^{\infty}(-1)^n
n^2A_n^2\cos\left(\frac{n\pi}{3}\right)\, . \eeq

Using the same arguments as in the previous case, usually $f_1<0$. Then, if $P>0$ the corresponding vortex breather is unstable, while if $P<0$ the breather is linearly stable.

\section{A model for hexagonal dusty plasma crystals \label{modelDP}}

 Dusty plasma crystals form various configurations, the most common of which is a two-dimensional hexagonal lattice \cite{Morfill}; see Fig.  \ref{lattice1}.
 The typical inter-site spacing is of the order of (or slightly above) unity (in units of Debye length $\lambda_D$). Nearest neighbor interactions are therefore assumed in modelling dust crystals.
\begin{figure}[htbp]
\begin{center}
		\includegraphics[width=7cm]{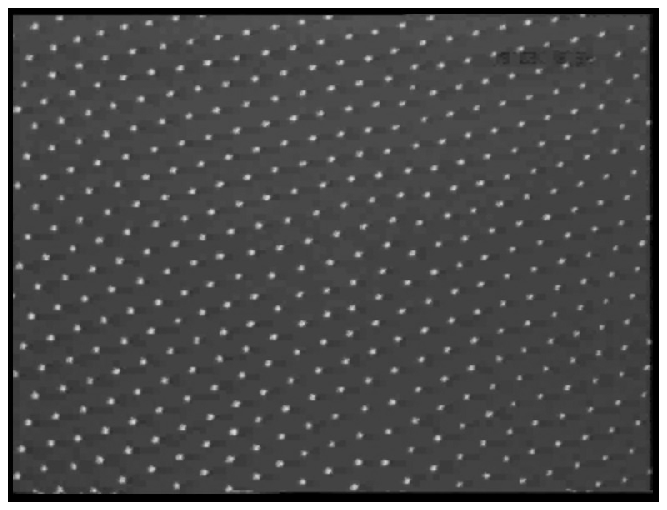}
	\caption{ A snapshot of the DP crystal by a CCD camera. Courtesy of Dmitry Samsonov (University of Liverpool, UK).}
	\label{lattice_camera}
	\end{center}
	\end{figure}

The vertical displacement $z_{nm}$ of the charged grain at site $(i, j)$ in a DP crystal obeys an equation of motion in the form:
\beq
\frac{d^2 z_{ij}}{dt^2} = \frac{1}{M} (F_e - M g) + F_{int, z} -\nu \frac{d z_{ij}}{dt}
\label{geneqmotion}
\eeq
We distinguish three contributions in the right-hand side (\emph{rhs}), to be specified below.

The first term corresponds to the the overall vertical force acting on a single grain (in the absence of coupling and damping); it consists of an upward force $F_e$ due to the electric field of the plasma sheath and of the force of gravity, $M g$, pointing downwards. At the levitated equilibrium position of the crystal, at height $z_0$, the two forces balance each other, viz. $F_e(z_0) - Mg = 0$. (We may set $z_0$ with no loss of generality, in the following.) Expanding the (variable) charge and electric field as $q(z) \approx q_0 + q_0' z + \frac{1}{2} q_0''(z) z^2 + ...$ and $E(z) \approx E_0 + E_0' z + \frac{1}{2} E_0''(z) z^2 + ...$ (the prime denotes differentiation with respect to $z$, while the index ``0" denotes evaluation at the equilibrium position), the overall vertical force reads
\begin{eqnarray}
F_e(z) - M g & \approx &(q_0 E_0' + q_0' E_0) z  \nonumber \\ &&  + \frac{1}{2} (q_0 E_0'' + 2 q_0' E_0' + q_0'' E_0) z^2 + ...  \nonumber \\
& \equiv &- M \omega_{0}^2 (z + \tilde a z^2 + \tilde b z^3 + ...) \nonumber \\
& =& - \frac{\partial V(z)}{\partial z}
\, ,
\label{Fe}
\end{eqnarray}
where the definition of all quantities is obvious. The transverse vibration eigenfrequency may be provided by a simple Bohm sheath theory \cite{VSC1, VSC2}. The anharmonic character of the vertical on-site sheath potential $V(z)$, which clearly appears in \textit{ab initio} calculations for low density and low pressure \cite{GS} and also in experiments \cite{Ivlev2000, Zafiu}, is measured by the coefficients $a$ and $b$, which can be directly inferred experimentally [see (\ref{anharmonicpot}) below; cf. Table 1]. The \emph{tilde} was used in the latter expression to remind that these coefficients here have dimensions, to be henceforth removed via an appropriate scaling, viz. $\tilde \alpha \rightarrow r_0 \alpha$, $\tilde \beta \rightarrow r_0^2 \beta$.

The vertical component $F_{int, z}$ of the total interaction force $\mathbf{F}_{int}$ (due to the electrostatic coupling to neighboring sites) consists of the sum of the six nearest neighbor contributions,
\begin{eqnarray}
F_{int} & \approx & - \frac{\partial U}{\partial r}\biggr|_{r=a} \sum_{l=1}^6 \frac{\delta z_l}{(a^2 + \delta z_l^2)^{1/2}} \nonumber \\  & \approx &
- \frac{\partial U}{\partial r}\biggr|_{r=a} \sum_{l=1}^6 \frac{\delta z_l}{a}
\equiv - M \omega_{T,0}^2 \sum_{l=1}^6 \frac{\delta z_l}{a}
\, ,
\label{Fint}
\end{eqnarray}
where $\delta z_l = z_l -z_{ij}$ (assumed $\ll a$) and the dummy index $l$ runs overs all six nearest neighbor -- to (i,j) -- indices in Figure 1.
For Debye interactions, viz. $U(r) = e^{-r/\lambda_D}/r$, the characteristic frequency $\omega_{T,0}$ is obtained as in (\ref{omegaT0Debye}).

The last term in the \emph{rhs} of (\ref{geneqmotion}) represents energy dissipation (damping), and involves the collision frequency $\nu$. This term will be neglected, since $\nu \ll \omega_0$ applies to number of real experiments \cite{Samsonov3}. The effect of damping will be included in future work.

It is straighforward to combine the latter two expressions, (\ref{Fe}) and (\ref{Fint}), into (\ref{geneqmotion}). Furthermore, one may scale space and time by the lattice spacing $a$ and the inverse eigenfrequency $\omega_0^{-1}$, to obtain a reduced (dimensionless) equation of motion.
One thus obtains an equation of motion in the form $M d^2 z_{ij}/dt^2 = - \partial H(p_{ij}, z_{ij})/\partial z_{ij}$, where the (classical) hamiltonian $H$ is precisely given by (\ref{ham}), setting $x_{ij} = z_{ij}$, $p_{ij} = M d z_{ij}/dt$ and $\w_p=1$ therein.  Note however that the coupling coefficient in (\ref{ham}) needs to be taken as negative-valued, viz. $\e = - \omega_{T, 0}^2/\omega_0^2$. The negative sign of $\e$, here arising naturally as an outcome of the algebra, was expected from the inverse dispersive character of vertical dust-lattice vibrations (see the negative slope in the observed dispersion curve, e.g., in Fig.  3 in \cite{Qiao}).

\section{Discrete breathers in a 2D dusty plasma crystals \label{resultsDP}}

Consider a Hamiltonian of the form (\ref{ham}) with $\e<0$ and
\beq
V(x) = x^2/2 + a x^3/3 + b x^4/4 \, .
\label{anharmonicpot}
\eeq
We shall consider three sets of typical values from real experiments \cite{melzer} for the nonlinearity parameters $a$, $b$ and for the coupling constant $\e$, shown in Table \ref{parameters}.
\begin{table}[!ht]
\begin{tabular}{|c|c|c|c|}
\hline
&$a$&$b$&$\e$\\
\hline
set I	& 0.01 &	-0.04 &	-0.034\\
\hline
set II	& 0.01 &	-0.06 &	-0.065\\
\hline
set III	& -0.21 &	-0.02 &	-0.17\\
\hline
\end{tabular}
\caption{Potential form coefficients occurring in laboratory experiments \cite{melzer}.}
\label{parameters}
\end{table}

We have tested the numerical stability of the solutions obtained above, for the values of Table \ref{parameters}. Our numerical analysis is two-fold. It first focuses on the dynamics of the Floquet multipliers of the corresponding DB, for given values of $a$, $b$ and $\e$, varying the latter in order to span a range of different values and thus identifying existing (in)stability regimes, in terms of the strength of the inter-grain electrostatic coupling. Once the stable/unstable regions were traced, a numerical simulation has been used to confirm our predictions for the breather stability. The main points of the analysis are presented below.

\subsection{Single site breathers}

\begin{figure}[!ht]
\begin{centering}
\begin{tabular}{cc}
\includegraphics[height=5cm]{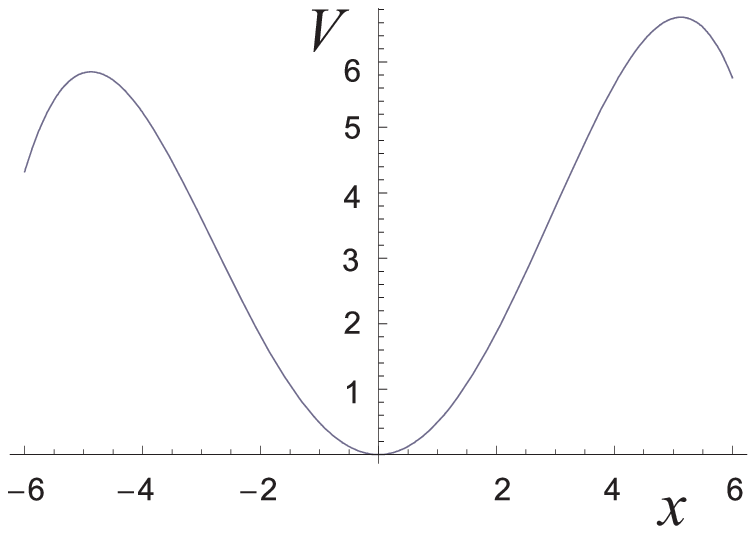}&\includegraphics[height=5cm]{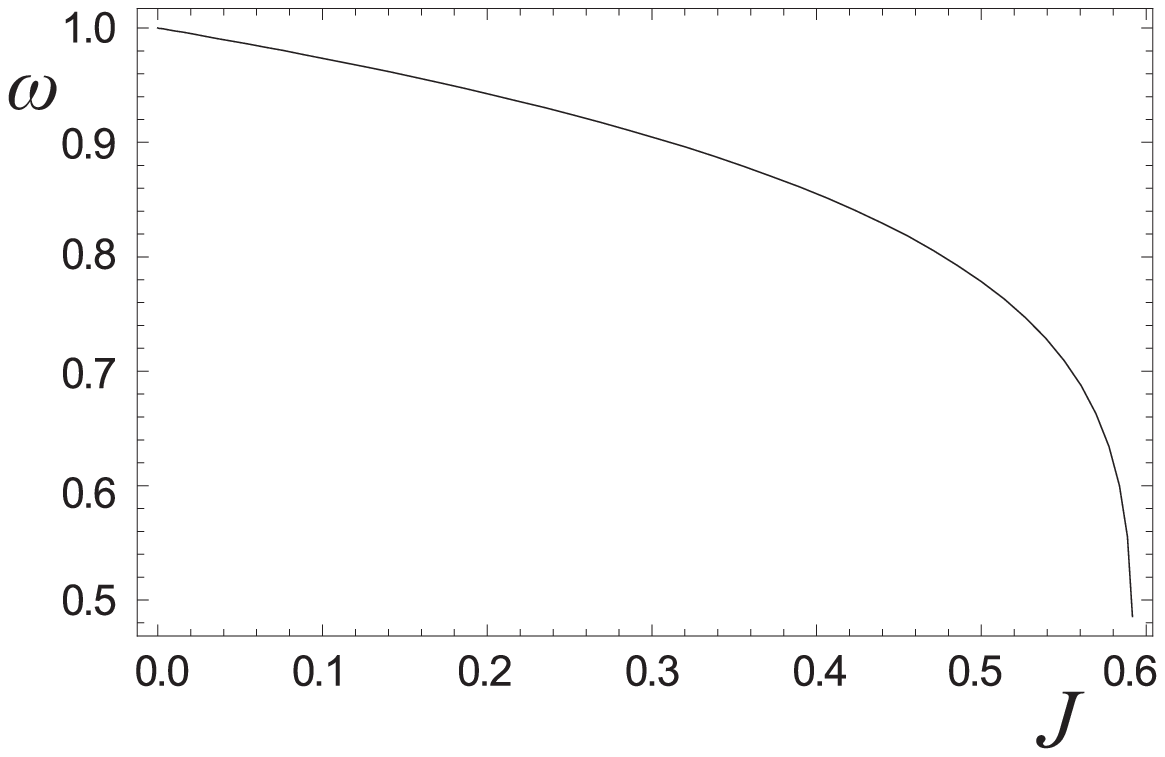}\\
(a)&(b)
\end{tabular}
    \caption{(Color online) (a) The on-site potential of the lattice for set I of Table \ref{parameters}, and (b) the frequency $\w$ of the oscillation for the same potential with respect to $J$.}
    \label{potential}
    \end{centering}
\end{figure}

\begin{figure}[!ht]
\includegraphics[width=5cm]{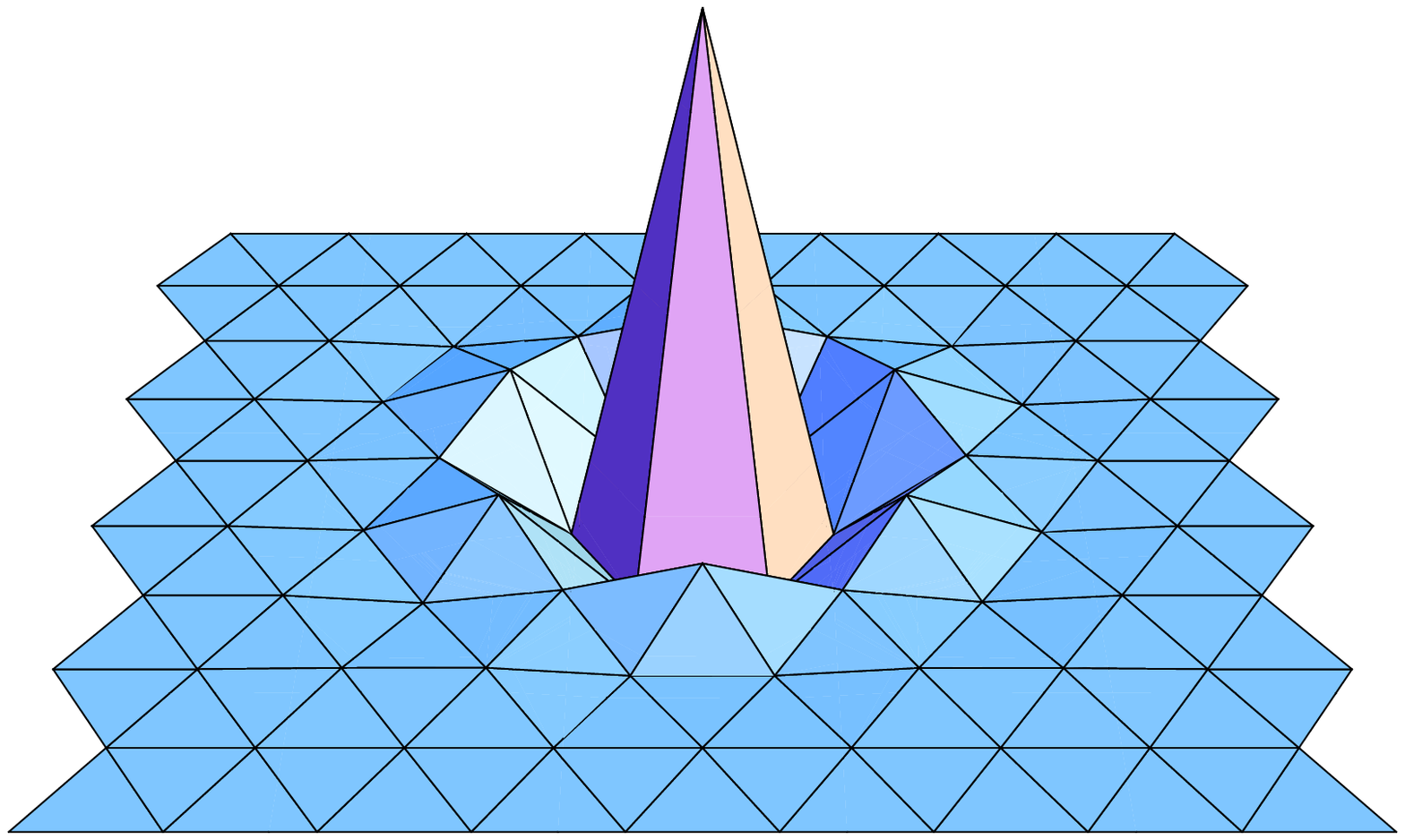}\includegraphics[width=5cm]{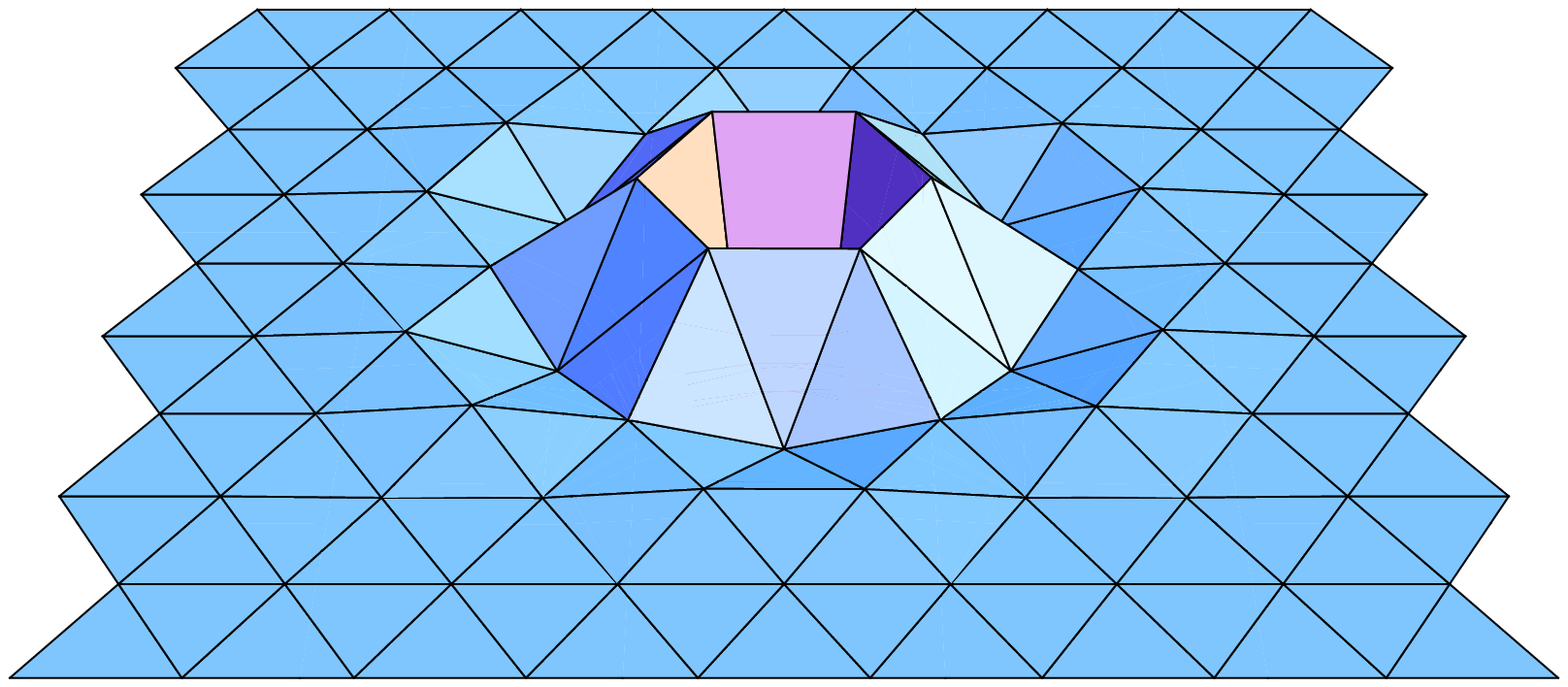}
\vspace{-1.2cm}
\caption{(Color online) A single site breather for $|\e|=0.035$. For a better representation, please refer to the video files provided in \cite{video_link}.}
\label{single}
\end{figure}

\begin{figure}[!ht]
\begin{centering}
\begin{tabular}{cccc}
\includegraphics[height=4cm]{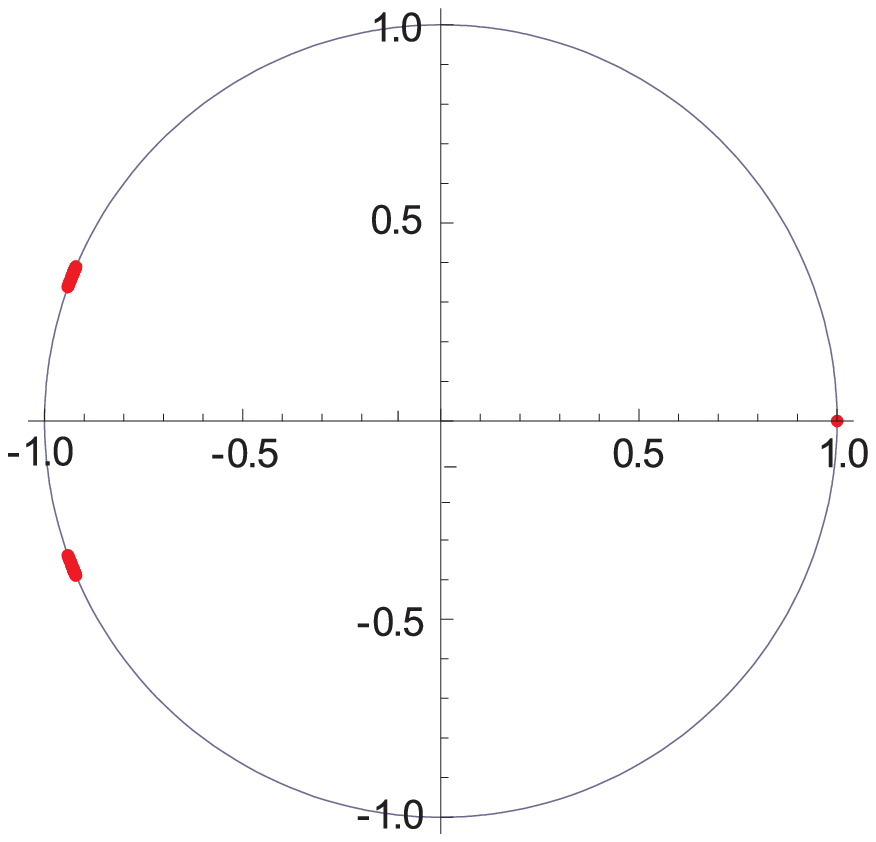}&\includegraphics[height=4cm]{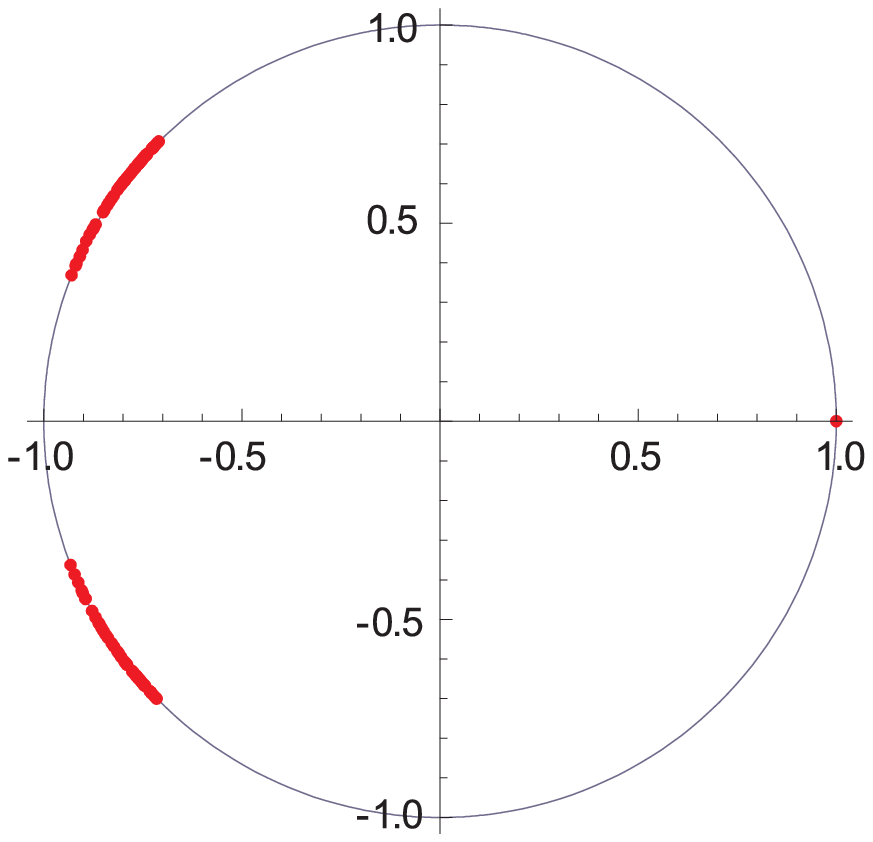}&\includegraphics[height=4cm]{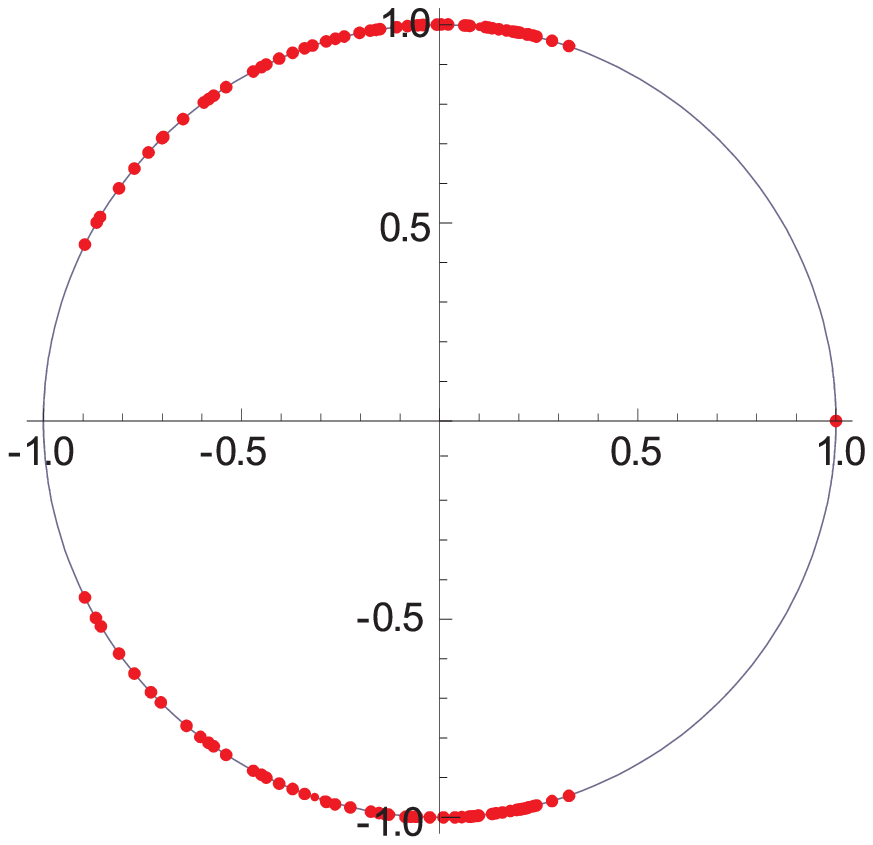}&\includegraphics[height=4cm]{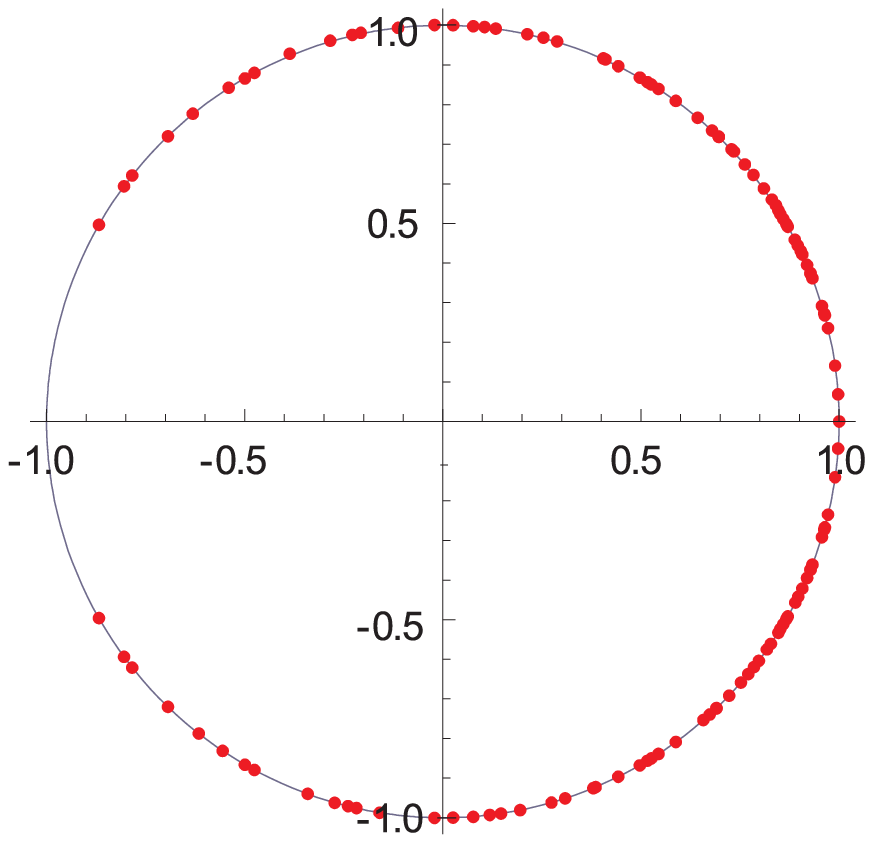}\\
$|\e|=0.001$&$|\e|=0.010$&$|\e|=0.035$&$|\e|=0.056$
\end{tabular}
    \caption{(Color online) The distribution of the Floquet multipliers corresponding to the breather depicted in Fig. \ref{single} for increasing values of $|\e|$.}
    \label{e_single}
    \end{centering}
\end{figure}

We will first examine a system with the first set of values in Table \ref{parameters}. The on-site potential of the lattice has the form of Fig.  \ref{potential}(a) and the corresponding frequency range of an individual oscillator $\w(J)$ is shown in Fig.  \ref{potential}(b). In order to compute a breather we have to consider one of the oscillators of the lattice to oscillate with a frequency $\w_b\neq m\w_p$. Note that in the specific case $\w_p=1$ due to normalization we have already performed. By taking a look in Fig.  \ref{potential}(b) we can very easily choose a frequency $\w_b$ which does not resonate with the phonon frequency $\w_p$. Although the graph Fig.  \ref{potential}(b) is drawn with respect to action $J$ there is a one-on-one correspondence with the amplitude of the oscillation which can be very easily computed even if we do not know the explicit form of the action-angle canonical transformation \cite{kouk}. For example the results presented in Fig.  \ref{e_single} are for $x_{max}=3.3$ which results to $\w_b\simeq0.692$.

We can see that for $|\e|=0.001$ we have a pair of multipliers at unity while the rest lie in two very small conjugate bundles. For increasing values of $|\e|$ these bundles (which consist the phonon band) spread along the unit circle and finally for $|\e|\simeq0.056$ they reach $+1$. This means that the frequency of the breather has reached the phonon band, so the breather can no longer exist. So, we can expect the existence of single site breathers in such a system since the value of $\e$ in set I is $|\e|=0.034$. Note that, in an infinite length lattice the phonon band would be continuous. In figs.\ref{e_single} it appears discrete because of the finite size of the system we use for computations. By using similar arguments we find that for sets II and III we can expect breather motions only up to $|\e|\simeq0.049$ and $|\e|\simeq0.037$ respectively. This is somehow anticipated, since higher values of $\e$ the continuous behavior of the system.

\subsection{Three-site breathers}

\begin{figure}[!ht]
\begin{tabular}{cc}
\includegraphics[height=5cm]{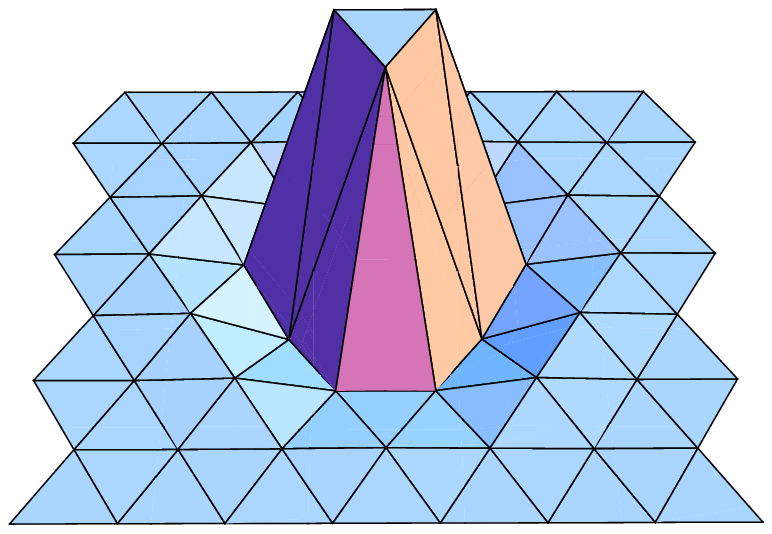}&\includegraphics[height=5cm]{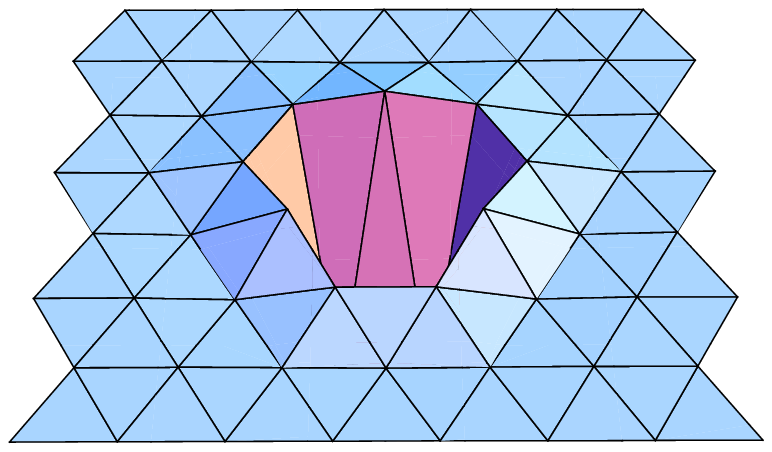}\\[-30pt]
(a)&(b)
\end{tabular}
\caption{(Color online) An in-phase three-site breather for $|\e|=0.015$.  For a better representation, please refer to the video files provided in \cite{video_link}.}
\label{in}
\end{figure}

We will consider again set I for our study. Due to Eq.\ref{s_in} and since  $\frac{\pa \w}{\pa J}<0$ (Fig.  \ref{potential}(b)) determines the linearly stable solution, at least for small values of $\e$, is the in-phase configuration (Fig.  \ref{in}).

\begin{figure}[htbp]
\begin{tabular}{cccc}
\includegraphics[height=4cm]{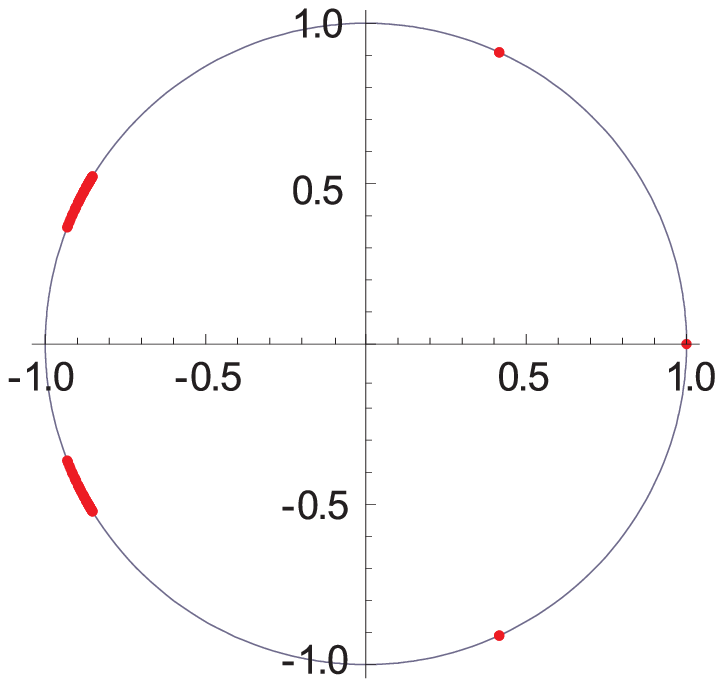}&\includegraphics[height=4cm]{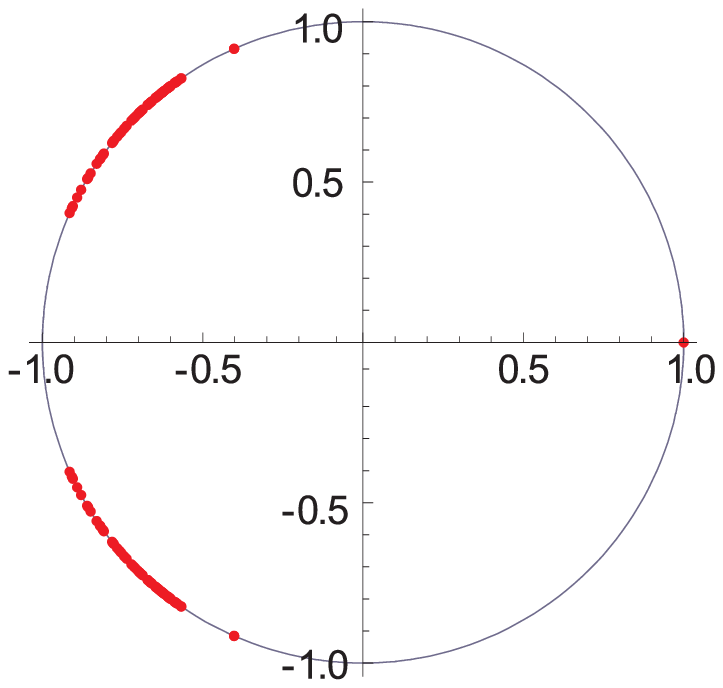}&\includegraphics[height=4cm]{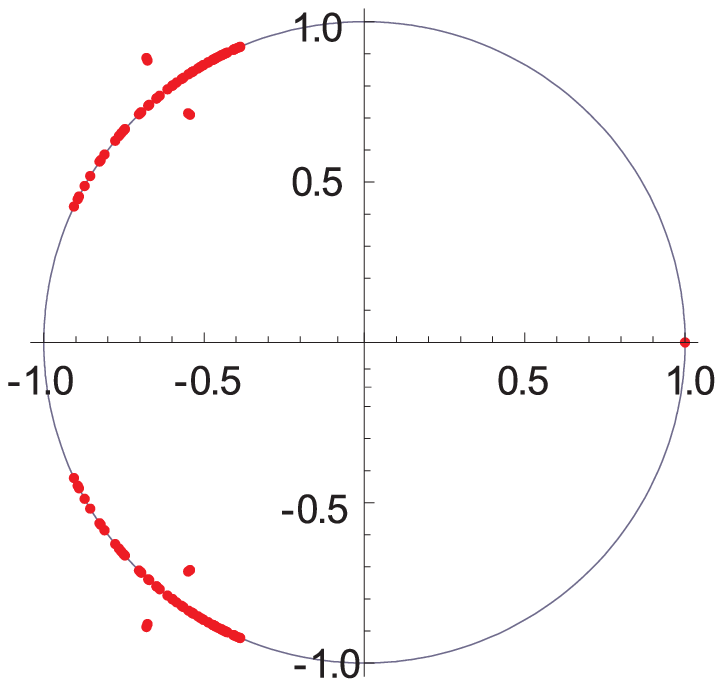}&\includegraphics[height=4cm]{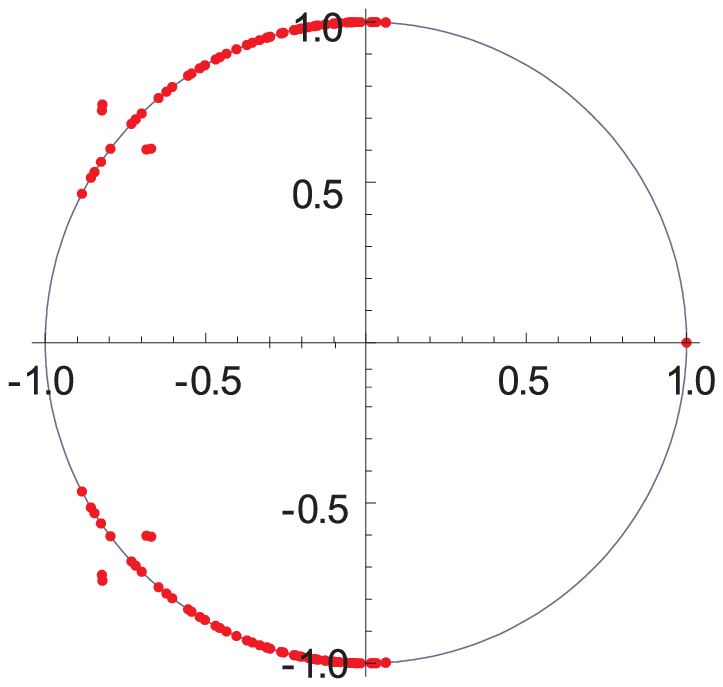}\\
$|\e|=0.005$&$|\e|=0.015$&$|\e|=0.020$&$\hspace{1cm}|\e|=0.030$
\end{tabular}
    \caption{(Color online) The distribution of Floquet multipliers for increasing $|\e|$ fot the breather depicted in Fig. \ref{in}. The last two images show the destabilization of the breather for $|\e|>0.017.$}
    \label{e_in}
\end{figure}

As it can be seen in Fig.  \ref{e_in}, the in-phase breather remains linearly stable until $|\e|\simeq0.017$ when a Hopf bifurcation occurs and it destabilizes (Fig.  \ref{e_in}), which means that after approximately 120 periods of oscillation the energy starts flowing to the rest of the lattice.

Following the same kind of arguments we conclude that for the sets II and III of Table \ref{parameters} the corresponding in-phase breathers remains stable until $|\e|\simeq0.016$ $|\e|\simeq0.013$ respectively.

\begin{figure}[!ht]
\begin{tabular}{cc}
\includegraphics[height=5cm]{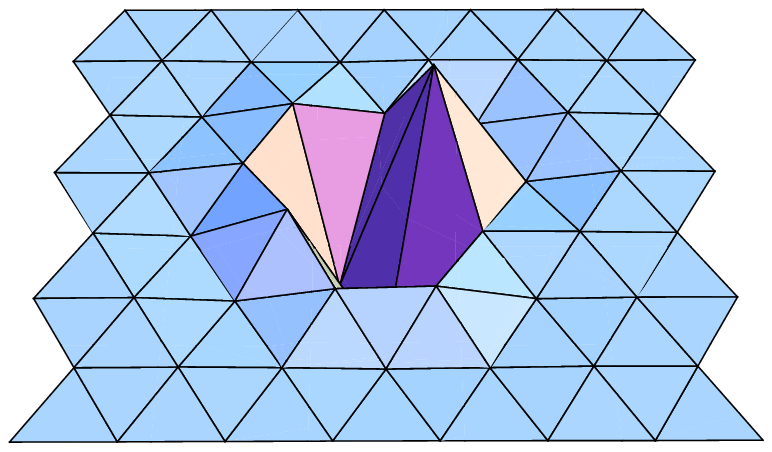}&\includegraphics[height=5cm]{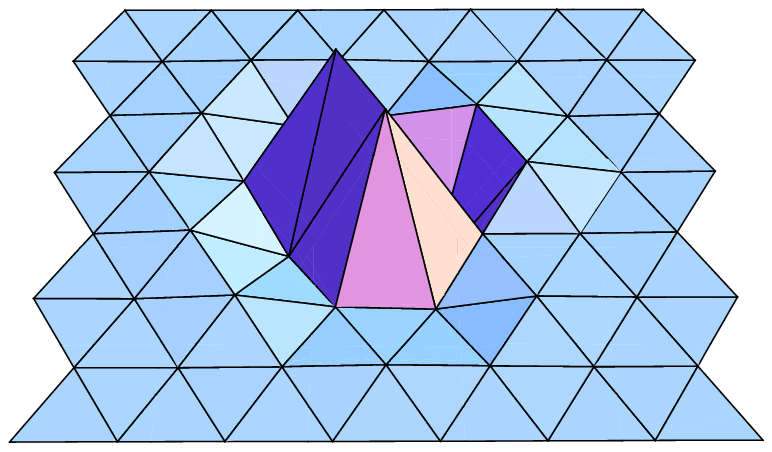}\\[-30pt]
(a)&(b)
\end{tabular}
\caption{(Color online) An out of phase three-site breather for $\e=0.005$. For a better representation, please refer to the video files provided in \cite{video_link}.}
\label{out}
\end{figure}

\begin{figure}[!ht]
\begin{tabular}{cccc}
\includegraphics[height=4cm]{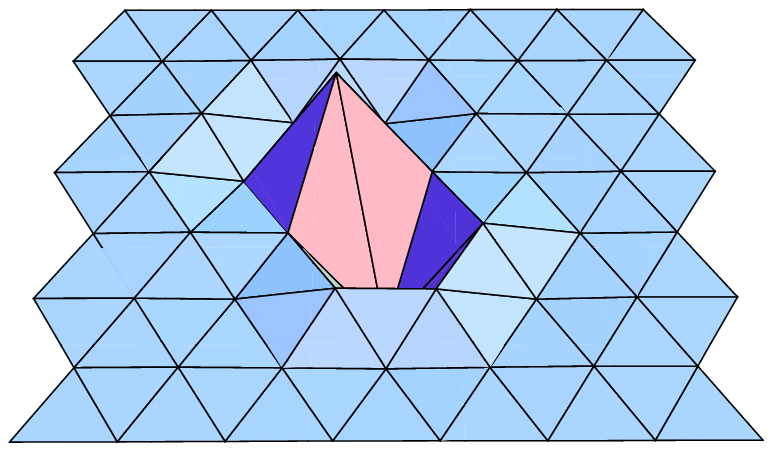}&\includegraphics[height=4cm]{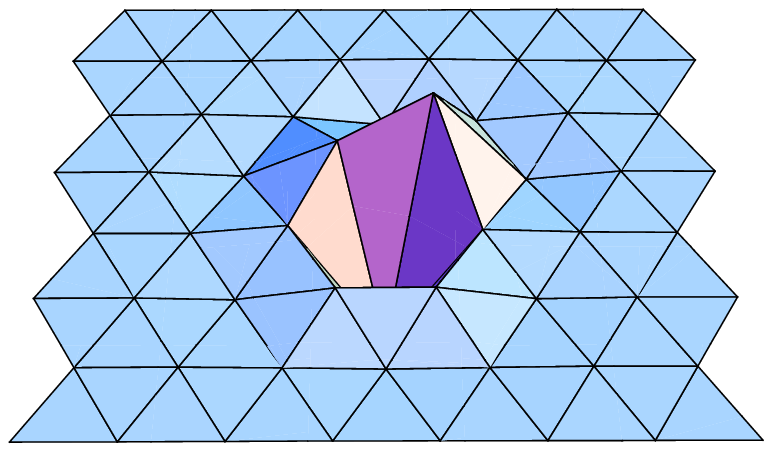}&\includegraphics[height=4cm]{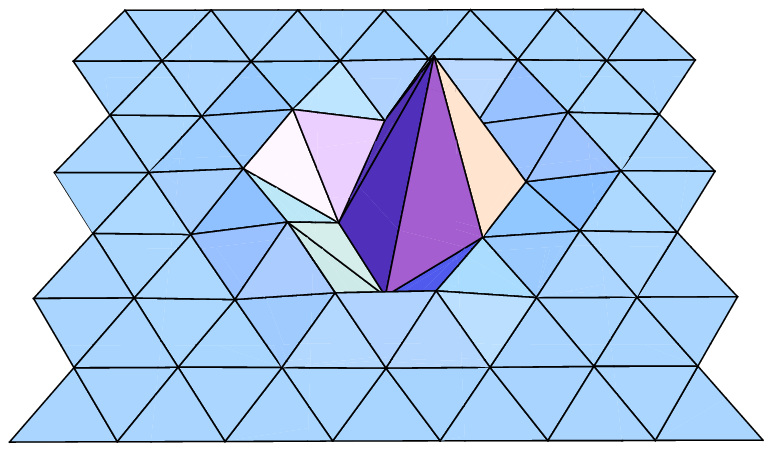}&\includegraphics[height=4cm]{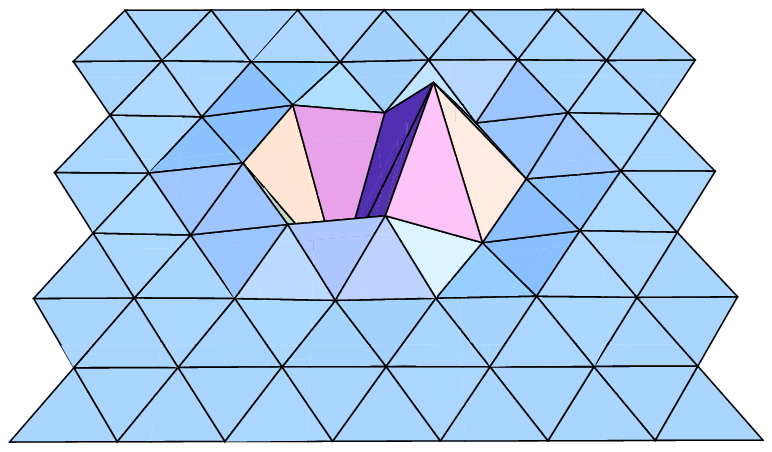}\\[-27pt]
(a)&(b)&(c)&(d)\\[-20pt] \includegraphics[height=4cm]{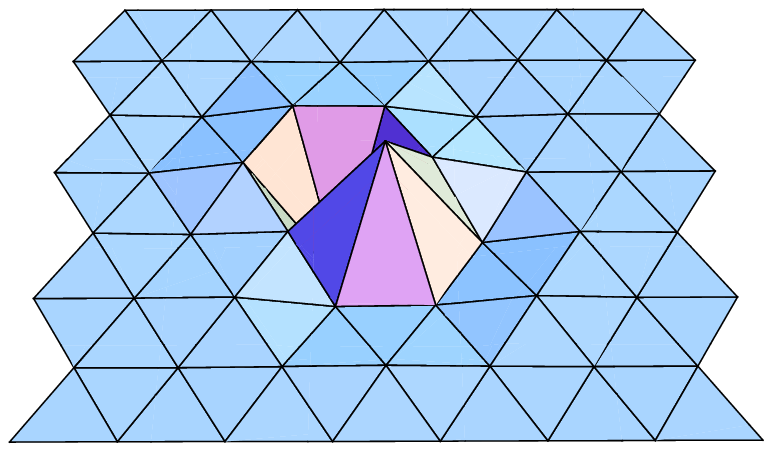}&\includegraphics[height=4cm]{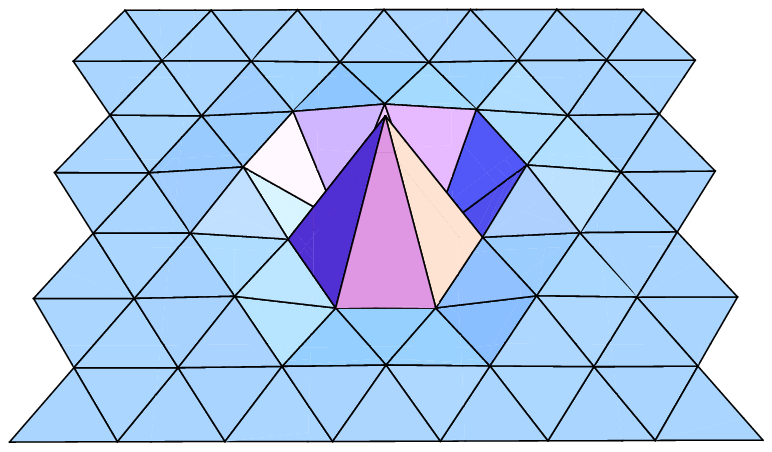}&\includegraphics[height=4cm]{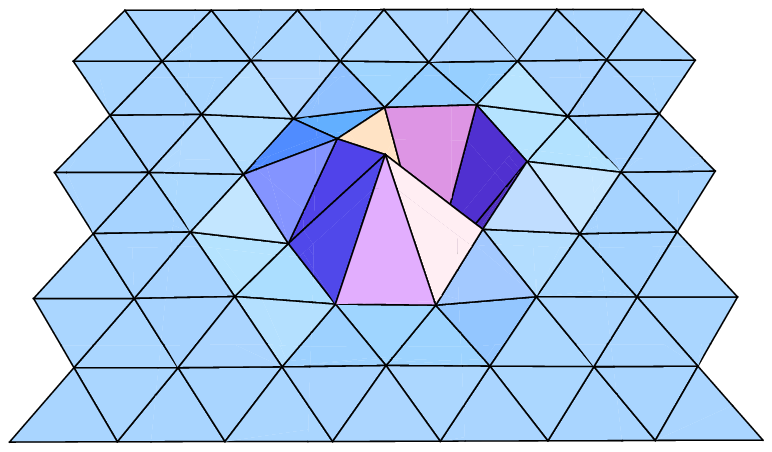}&\includegraphics[height=4cm]{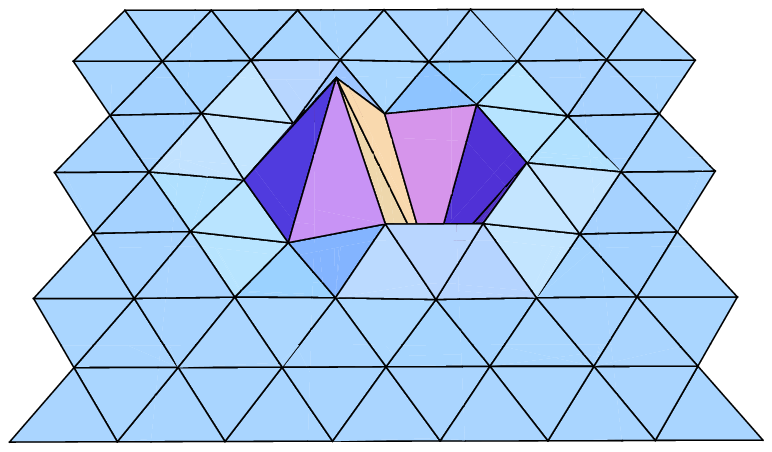}\\[-27pt]
(e)&(f)&(g)&(h)
\end{tabular}
\caption{(Color online) A vortex breather for $\e=0.008$.  For a better representation, please refer to the video files provided in \cite{video_link}.}
\label{vortex}
\end{figure}

\begin{figure}[!ht]
\begin{tabular}{cc}
\includegraphics[height=5cm]{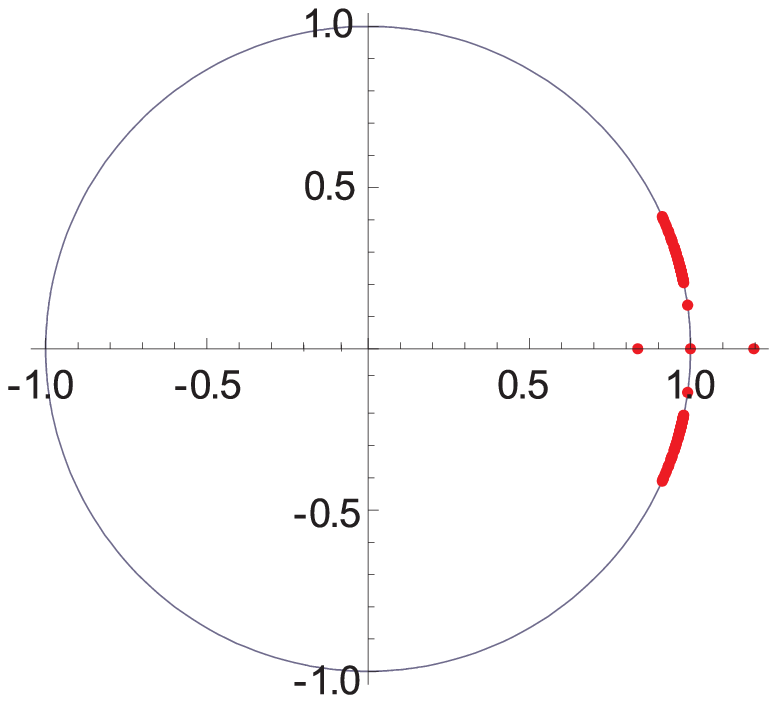}&\hspace{0.5cm}\includegraphics[height=5cm]{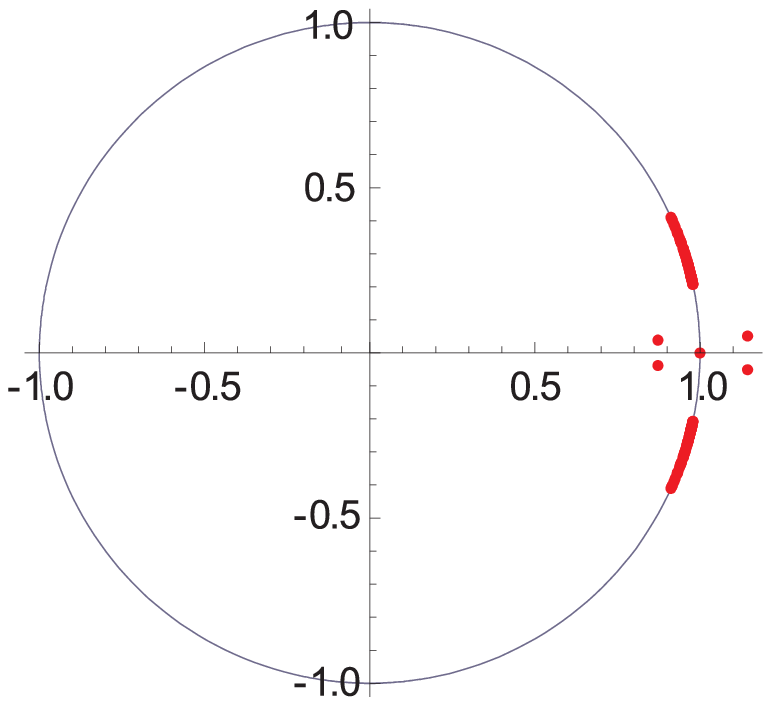}\\
(a)&\hspace{0.5cm}(b)
\end{tabular}
\caption{(Color online) Floquet multipliers distribution for (a) the out of phase three-site breather depicted in Fig. \ref{out} and (b) the vortex breather depicted in Fig. \ref{vortex}, both for $\e=0.008$. Note that the both have multipliers outside the unit circle, even for small values of $\e$, which imply instability.}
\label{e_out_vortex}
\end{figure}

The out of phase (Fig.  \ref{out}) and vortex (Fig.  \ref{vortex}) configurations have [due to Eqs. (\ref{s_out}), (\ref{s_vortex})] one and two pairs of characteristic exponent which have real part to leading order of approximation, so they are linearly unstable for $|\e|$ arbitrary small. In Fig.  \ref{e_out_vortex}, in can be seen that for $|\e|>0$ the out of phase breather possesses a pair of multipliers in the real axis while the vortex breather possesses a complex quadruple.

\section{Conclusions}

In this article, we have established the occurrence in a hexagonal dusty plasma lattice of localized vibrational modes, in the form single-site or three-site breathers. Since transverse dust-lattice vibrations in a dusty plasma lattice are characterized by an inverse dispersion (backward wave),
we have identified a phenomenological Klein-Gordon type Hamiltonian with negative coupling $\e$ and a quartic polynomial on-site potential to take into account this behavior. The parameters of the potential are given from real experiments and are divided into three sets originating from three different experimental configurations.
We have considered the first set of values, since this best satisfies the discreteness requirement, having $\e=-0.034$. For this configuration we conclude that single-site breathers can be supported, but the three-site ones are unstable for this value of the coupling.
However, real values of the coupling parameter $\epsilon$ in experiments span a large region of values, so that three-site breathers are expected to be stable for a different parameter set.

\acknowledgments The work of IK was supported by a UK EPSRC Science and Innovation Award (Centre for Plasma Physics, Queen's University Belfast).
Part of the work was carried out by IK during a research visit to the University of Sydney.
IK is grateful to UK Royal Society for the award of a travel grant, and to University of Sydney for its hospitality and local support provided during that visit. VK would like to thank the Centre for Plasma Physics, Queen's University of Belfast for its kind hospitality.

\bibliographystyle{plain}

\end{document}